\documentclass[prl,twocolumn,showpacs,preprintnumbers,amsmath,amssymb]{revtex4}

\usepackage{amsfonts} 

\usepackage{hyperref}

\usepackage{color} 
 
\usepackage[utf8,applemac]{inputenc} 
\usepackage{tensor}
\usepackage{tikz}
\usepackage{graphicx}
\usepackage{subfig}	

\usetikzlibrary{positioning}
\usetikzlibrary{calc}

\newcommand{\bea}{\begin{eqnarray}}
\newcommand{\eea}{\end{eqnarray}}
\newcommand{\be}{\begin{equation}}
\newcommand{\ee}{\end{equation}}

\def\p{\partial}
\def\eps{\epsilon}

\usepackage{mathtools}

\begin{document}
\title{Infinite towers of supertranslation and superrotation memories}
  
  \author{Geoffrey Comp\`ere}
\affiliation{%
Physique Th\'eorique et Math\'ematique, \\ Universit\'e Libre de
    Bruxelles\\and\\ International Solvay Institutes\\ Campus
    Plaine C.P. 231, B-1050 Bruxelles, Belgium
}

\begin{abstract}
A framework that structures the gravitational memory effects and which is consistent with gravitational electric-magnetic duality is presented. A correspondence is described between memory observables, particular subleading residual gauge transformations, associated overleading gauge transformations and their canonical surface charges. It is shown that matter-induced transitions can generate infinite towers of independent memory effects at null infinity. These memories are associated with an infinite number of conservation laws at spatial infinity which lead to degenerate towers of subleading soft graviton theorems. 
 It is shown that the leading order mutually commuting supertranslations and (novel) superrotations are both associated with a leading displacement memory effect, which suggests the existence of new boundary conditions. 
\end{abstract}


\maketitle


Memory effects can be formulated as the difference between two observables defined in the initial and final non-radiative regions of spacetime after energy has escaped to null infinity. 
Distinct classes of memory effects have been found: the displacement memory effect \cite{Zeldovich:1974aa,Christodoulou:1991cr}, the velocity kick memory effect \cite{Braginsky:1986ia} and, more recently, the spin and center-of-mass memory effects \cite{Pasterski:2015tva,Nichols:2018qac,Himwich:2019qmj}. The aim of this Letter is to propose a framework for describing memory effects and illustrate some of its features. This construction will extend the relationship between memories, gauge transformations and soft theorems \cite{Strominger:2017zoo}. For definiteness, I will focus on 4-dimensional asymptotically flat spacetimes in general relativity coupled to matter. Retarded coordinates will be denoted as $x^\mu = \{ u,r,x^A \}$ where $x^A = \{\theta,\phi\}$ are angles on a 2-sphere.

\vspace{10pt}
{\noindent \bf The memory $\leftrightarrow$ gauge transformation map.} I consider non-local gauge-invariant observables $O=O(\theta,\phi)$ defined by integration of a functional of the gauge-invariant fields defined in the asymptotic null region $\mathcal I^+$ (i.e. the Weyl tensor for pure gravity)  between infinite past $\mathcal I^+_-$ and infinite future $\mathcal I^+_+$ retarded times. Such an observable $O$ is a memory observable if and only if it can be written as the difference between a \emph{local memory field} $\varphi_+$ defined at $\mathcal I^+_+$ and a \emph{local memory field} $\varphi_-$ defined at $\mathcal I^+_-$, $O = \varphi_+ - \varphi_-$. The fields $\varphi_\pm$ might not exist in arbitrary gauges. The definition only requires that one gauge exists where $\varphi_\pm$ can be defined. The observable $O$ is gauge-invariant and it vanishes if all gauge-invariant fields asymptotically vanish at null infinity.

Only field configurations that are non-radiative at future and past retarded times are considered. For field configurations where the gauge-invariant fields asymptote to zero at future and past retarded times, two residual gauge transformations have to exist that shift each local memory field to zero. Since the memory observable is gauge-invariant, the residual gauge transformation $\delta$ shifts identically $\varphi_-$ and  $\varphi_+$, $\delta \varphi_- = \delta \varphi_+$.  I assume that physical field configurations obey fall-off boundary conditions at infinity. It implies that $\varphi_\pm$ are subleading components of the fields and the associated gauge transformations $\delta$ are subleading. \emph{Memory observables are therefore in one-to-one correspondence with particular subleading residual gauge transformations in the gauge where the local memory fields are defined.} Boundary conditions on the physical fields restrict the possible gauge transformations that they asymptote to.

\vspace{10pt}
{\noindent \bf The subleading towers.}  Motivated by several independent results \cite{Kapec:2016jld,Campiglia:2016efb,Compere:2017wrj,Hamada:2018vrw,Himwich:2019qmj}, I will only consider memory observables whose local memory fields can be defined in harmonic gauge. The solutions to $\square \xi^\mu = 0$ in Minkowski spacetime form 2 scalar and 1 vector representations of $SO(3)$. Now, it exists for each representation an infinite tower of gauge transformations $\delta_{(N)}$ labelled by an integer $N \geq 0$. The explicit gauge transformations are given by
\bea
\xi^{T}_{(N)} &\stackrel{\mathclap{\tiny\mbox{$N > 0$}}}{=} & \frac{1}{r^{N}} \Big[ \left( T + O(r^{-2}) \right) \p_u + \Big( \frac{N(1-N)}{2r}T\nonumber \\ 
&&\hspace{-32pt} +O(r^{-2}) \Big) r\p_r +\Big( -\frac{1}{2}\nabla^A T+\frac{u}{4N r} (\Delta \nonumber\\
&&\hspace{-32pt}   +3N(N-1)-1)\nabla^A T + O(r^{-2})\Big) \frac{1}{r}\p_A \Big] ;\\
\xi^R_{(N)} &\stackrel{\mathclap{\tiny\mbox{$N > 0$}}}{=} & \frac{1}{r^{N}} \Big[O(r^{-2}) \p_u +  \left( -\frac{1}{r} \nabla \cdot R  +O(r^{-2})\right)r \p_r  \nonumber \\
&&\hspace{-32pt}+ \Big(  R^A + \frac{u}{2N r} (- \Delta R^A + 2 \nabla^A \nabla \cdot R  \nonumber \\
&&\hspace{-32pt}+ (1+N-N^2) R^A) + O(r^{-2})\Big)\frac{1}{r}\p_A \Big] ; \\
\xi^{W}_{(N)} &\stackrel{\mathclap{\tiny\mbox{$N > 1$}}}{=} &\frac{1}{r^{N}} \Big[   \Big( -\frac{u}{N-1}W +\frac{u^2}{(N-1)r} (\frac{\Delta}{N} +\frac{N-2}{2}   )W \nonumber \\
&& \hspace{-32pt}+ O(r^{-2}) \Big) \p_u + \Big( W  -\frac{u}{2(N-1)r} (\Delta+N(N-3))W+ \nonumber 
\eea
\bea
&& \hspace{-32pt}  O(r^{-2}) \Big) r\p_r +\Big( -\frac{u}{N-1}\nabla^A W+ \frac{u^2}{2(N-1)r} \nonumber\\
&&\hspace{-32pt}   (\frac{\Delta}{N}+N-2)\nabla^A W+ O(r^{-2})\Big) \frac{1}{r}\p_A \Big];\label{genxi}\\
\xi^{T}_{(0)} &=& \left( T + O(\frac{\log r}{r^2}) \right)\p_u + \left( \frac{\Delta T}{2r}  + O(\frac{\log r}{r^2}) \right)r\p_r \nonumber\\
&&\hspace{-30pt}+ \left(-\nabla^A T +O(r^{-1}) \right)\frac{1}{r}\p_A; \\
\xi^R_{(0)} &=& O(r^{-2}) \p_u +\Big( - \frac{\nabla \cdot R}{r}-\frac{u \log r}{2r^2} \Delta \nabla \cdot R \nonumber \\
&& \hspace{-30pt} +O(r^{-2}) \Big) r \p_r  +\Big( R^A + \frac{u \log r}{2r} (\Delta R^A - 2 \nabla^A \nabla \cdot R\nonumber \\
&& \hspace{-30pt}  - R^A)  +O(r^{-2})\Big)\frac{1}{r}\p_A;\\ 
\xi^{W}_{(1)} &=& \left( \frac{u \log r}{r}W + O(\frac{\log r}{r^2}) \right)\p_u \nonumber \\
&& \hspace{-30pt} + \left( \frac{W}{r} + O(\frac{\log r}{r^2}) \right)r\p_r \nonumber\\
&&\hspace{-30pt} + \left( \frac{u \log r}{r}\nabla^A W +O(\frac{\log r}{r^{2}}) \right)\frac{1}{r}\p_A; \\
\xi^{W}_{(0)} &=&\left(uW + \frac{u^2 \log r}{r} \Delta W + O(r^{-1}) \right)\p_u \nonumber \\ 
&&\hspace{-30pt}+ \left( W +\frac{u}{2r}\Delta W+ O(r^{-1}) \right)r\p_r \nonumber \\
&&\hspace{-30pt} + \left( u \nabla^A W +\frac{u^2\log r}{2r}\Delta \nabla^A W+O(r^{-1}) \right)\frac{1}{r}\p_A.
\eea

The generators $\{\xi_{(N)}^{T}, \xi_{(N)}^{W}, \xi_{(N)}^{R} \}$  depend upon arbitrary functions $T=T(\theta,\phi)$, $W=W(\theta,\phi)$ and $R^A=R^A(\theta,\phi)$. In short, the supertranslation and superrotation towers are generated by
\bea
\xi^T_{(N)} &=& r^{-N} (T \p_u -\frac{1+\delta_{N,0}}{2r} \nabla^A T \p_A )+\dots ,\label{xiT} \\
\xi^R_{(N)} &=& r^{-N} R^A \frac{1}{r}\p_A + \dots \label{xiR}
\eea
Some cases are familiar:  $\xi_{(0)}^{T}$ are the BMS asymptotic supertranslation symmetries and  $\xi_{(1)}^{R} $ are the subleading Diff$(S^2)$ transformations found in \cite{Himwich:2019qmj}. The \emph{novel leading superrotations} $\xi_{(0)}^{R}$ are on the same footing as the supertranslations since asymptotically $\p_u \sim r^{-1} \p_A$. They are $1/r$ subleading with respect to the Lorentz transformations. For generic $N$, one can map a subleading gauge transformation $\delta_{(N)}$ to an overleading gauge transformation $\delta_{(-N)}$, which transforms the metric at leading order or above, by analytic continuation $N \mapsto -N$. Examples of overleading transformations   are the superrotations defined in \cite{Barnich:2009se,Campiglia:2014yka} which are combinations of  $\xi^R_{(-1)}$ and $\xi^{T=\nabla \cdot R}_{(-1)}$. Finally, the generators $\xi_{(N)}^{W}$ are Weyl transformations given for $N=0$ by
\bea
\xi^W_{(0)} &=& W (r \p_r + u \p_u ) + u \nabla^A W \frac{1}{r}\p_A + \dots\label{xiW}
\eea
The vectors form an algebra which is determined by their leading order commutator. The supertranslations $\xi_{(0)}^{T}$ and the superrotations $ \xi_{(0)}^{R}$ commute.

\vspace{10pt}
{\noindent \bf Gravitational electric-magnetic duality.} The set of vectors $N \geq 1$ is complete under a notion of gravitational electric-magnetic duality\footnote{Gravitational duality for $N=0$ generates unphysical NUT charges and is therefore not enforced.}. The structure of electric-magnetic duality is richer in gravity as compared to electromagnetism thanks to the property that the gauge parameters themselves can be dualized, which dispense with the need for a dual potential formulation to describe all relevant canonical charges \cite{Compere:2011db,Compere:2017wrj}. The electric-magnetic dual of an infinitesimal gauge transformation $\xi^\mu$ is defined as $(\star \xi)^\mu \equiv \eps^{\mu\alpha\beta\gamma}n_\alpha \p_\beta \xi_\gamma$. Here $n^\mu \p_\mu =r \p_r + u \p_u$ is normal to $dS_3$ slices of Minkowski spacetime outside the lightcone at the origin. It is also the zero mode of the Weyl transformation $\xi_{(0)}^{W}$. It can be checked that the duality acts as 
\bea
\star \xi^{T}_{(N)} &=& \frac{N+1}{2} \xi^{R=\ast \nabla T}_{(N)}, \nonumber \\
\star \xi^R_{(N)} &=& (1-N) \xi^{\ast R}_{(N)} - N  \xi^{W = \nabla \cdot (\ast R)}_{(N+1)},\label{dual}\\
  \star \xi^{W}_{(N)} &=& 0 , \nonumber
\eea
for all $N \geq 1$, which proves the claim. Here $(\ast \nabla T)^A = \eps^{AB} \p_B T$, $(\ast R)^A \equiv \gamma^{AB}\eps_{BC} R^C$, $ \nabla \cdot (\ast R) = \eps^{AB} \p_A R_B$ where $\eps_{AB}$ is the volume form and $\gamma_{AB}$ the unit metric on the sphere (used to lower and raise indices).

\vspace{10pt}
{\noindent \bf Matter-induced transitions.} The transition between the initial and the final non-radiative regions depends upon the transition of all local memory fields $\varphi_+ - \varphi_-$. A natural question is whether or not each memory observable associated with each residual transformation listed above (i.e. generated by $\xi^{T}_{(N)}$, $\xi^{R}_{(N)}$, $\xi^{W}_{(N)}$) can be independently generated by physical processes. As a proof of principle, I consider impulsive transitions between an initial vacuum and a final vacuum, as considered by Penrose \cite{Penrose:1972aa,Nutku:1992aa}. The metric of the impulsive vacuum transition associated with a supertranslation is given by the shockwave $\eta_{\mu\nu}+ \Theta(u) \mathcal L_{\xi^T_{(N)}}\eta_{\mu\nu}$ where $\Theta(u)$ is the step function and $\eta_{\mu\nu}$ is the Minkowski metric. The Einstein tensor can be obtained by simple algebra. One then deduces from Einstein's equations that the following matter stress-tensor generates such a transition: 
\begin{align}
&T^{uu}\sim T^{ur}\sim T^{rr}\sim r^{-N-1},\quad T^{rA} \sim r^{-N-2},\nonumber \\
&T^{uA} \sim r^{-N-3},\quad T^{AB} \sim r^{-N-4}.
\end{align}
One can repeat the exercise for the impulsive vacuum transition associated with superrotations. The metric is $\eta_{\mu\nu}+ \Theta(u) \mathcal L_{\xi^R_{(N)}}\eta_{\mu\nu}$ and the required matter stress-tensor is 
\begin{align}
&T^{rr}\sim r^{-N-1},\quad T^{uu} \sim T^{ur} \sim T^{rA} \sim r^{-N-2},\nonumber \\
&T^{uA} \sim r^{-N-3},\quad T^{AB} \sim r^{-N-4}.
\end{align}
Finally, for Weyl transformations one finds
\begin{align}
&T^{uu}\sim T^{ur}\sim T^{rr}\sim r^{-N-1},\quad T^{rA} \sim r^{-N-2},\nonumber \\
&T^{uA} \sim T^{AB} \sim r^{-N-3}.
\end{align}
Since such behavior is independent from one another for each of the 3 choices and each $N$, I deduce that suitable matter can generate each transition independently. Note that if matter is generated by specific fields, the multipole structure of the fields will constraint which possible memory effects can occur in that theory.

\vspace{10pt}
{\noindent \bf The memory observables.} What are the observables? The metric perturbation asymptote to $\delta_{(N)} g_{\mu\nu}$ at early and late retarded times, which allows by construction to identify the memory fields in relevant components $\varphi_\pm = \{ C^u_{(N)}(\theta,\phi), C^r_{(N)}(\theta,\phi),C^A_{(N)}(\theta,\phi)\}$, which are associated with $\xi^T_{(N)}$, $\xi^W_{(N)}$ and $\xi^R_{(N)}$ at $\mathcal I^+_+$ and $\mathcal I^+_-$, respectively. The two scalar observables are derived quantities in terms of 
\bea
\Delta^{u+}_{(N)} = C^u_{(N)} \Big|^{\mathcal I^+_+}_{\mathcal I^+_-} ,\qquad \Delta^{r+}_{(N)} = C^r_{(N)} \Big|^{\mathcal I^+_+}_{\mathcal I^+_-}.
\eea
A derived observable of $\Delta^{u+}_{(0)}$ is the displacement memory effect described in \cite{Strominger:2014pwa}. 

Vectorial memory observables can be defined as a retarded time delays $\Delta^+ u$, $\Delta_{\star}^+ u$ defined for 2 counter-propagating null rays along a ring $R$ of circumference $2\pi L$ defined in the asymptotic future null region, 
\bea
\Delta^+ u &=& \frac{1}{2 \pi L} \oint_R \gamma_{AB} C^B_{(N)} dx^A \Big|^{\mathcal I^+_+}_{\mathcal I^+_-};\\
\Delta_\star^+ u &=& \frac{1}{2 \pi L} \oint_R \eps_{AB} C^{B}_{(N)} dx^A \Big|^{\mathcal I^+_+}_{\mathcal I^+_-}.
\eea
For $N=1$ the first observable exactly reduces to the spin memory \cite{Pasterski:2015tva,Himwich:2019qmj}. I expect that the second observable for $N=1$ relates to the center-of-mass memory \cite{Nichols:2018qac}. All other memory effects are new, but many may become trivial depending on the radiative boundary conditions.

A more straightforward memory observable associated to each supertranslation and superrotation is a displacement memory effect at the corresponding subleading order. I consider again the matter-induced vacuum transition in the impulsive limit, as discussed above. The shockwave that encodes such a transition is $\eta_{\mu\nu}+ \Theta(u) \mathcal L_{\xi_{(N)}}\eta_{\mu\nu}$ where $\xi_{(N)}$ is either $\xi^T_{(N)}$ or $\xi^R_{(N)}$. In either case, its Riemann tensor is 
\bea
R_{u A u B} \sim r^{-N-1} \p_u^2 \Theta(u). \label{eq1}
\eea
I consider two observers with 4-velocity given by $\p_u$ (plus subleading corrections that are required for a consistent normalization), with vanishing radial deviation $s^r = 0$ and with angular deviation $s^A$.  The geodesic deviation equation implies  
\bea
r^2 \gamma_{AB}\p_u^2 s^B = R_{uA u B} s^B. \label{eq2}
\eea
Substituting \eqref{eq1} into \eqref{eq2} leads to a $N$-th order subleading displacement memory effect. In particular, the superrotations $\xi^R_{(0)}$ are associated with a displacement memory effect at the same order as the one associated with supertranslations, i.e. a leading displacement memory effect. 

\vspace{10pt}
{\noindent \bf Multipole symmetries.} Multipole moments are Noether charges associated with multipole symmetries \cite{Compere:2017wrj}. Such multipole symmetries are in fact particular linear combinations of leading and overleading supertranslations, superrotations and Weyl transformations as I now show. The mass multipole symmetries $K_\chi$, the current multipole symmetries $L_\chi$ and the momentum multipole symmetries $P_\chi$ as defined in \cite{Compere:2017wrj} are
\bea
K_\chi &=& \chi \p_u +(u+r) P_\chi , \\   
L_\chi &=& (*\nabla \chi)^A \frac{1}{r}\p_A =  \eps^{AB} \p_B\chi \frac{1}{r}\p_A,\\
P_\chi &=& \vec{\nabla} \chi \cdot \vec{\p}  = \p_r \chi (\p_r - \p_u) + \frac{1}{r^2} \nabla^A \chi \p_A,
\eea
where $\chi$ is harmonic. For $\chi = r^N Y_{N,m}(\theta,\phi)$ where $Y_{N,m}(\theta,\phi)$ is a standard spherical harmonic, one has
\bea
K_\chi &=& (1-N) \xi^{T=Y_{N,m}}_{(-N)}+\frac{3-N}{2}\xi^{R=\nabla Y_{N,m}}_{(-N)} \nonumber\\&& 
-N^2 \xi^{W=Y_{N,m}}_{(1-N)} ,\\
L_\chi &=& \xi^{R= * \nabla Y_{N,m}}_{(-N)} ,\\
P_\chi &=& -N \xi^{T=Y_{N,m}}_{(1-N)}+\frac{2-N}{2} \xi^{R=\nabla Y_{N,m}}_{(1-N)}.
\eea
Moreover, the duality \eqref{dual} implies  
\bea
\star P_\chi &=& 0,\qquad \star K_\chi = 2 L_\chi, \\
\star L_\chi &=& (N+1) \left( (N-1) \chi \p_u - K_\chi \right),
\eea
which extends the dualities found in \cite{Compere:2011db,Compere:2017wrj}.  

In the second part of this Letter, I will describe how such towers of memories are associated to towers of subleading soft theorems following a sequence of equivalences:
\begin{figure}[!h]\vspace{-10pt}
\centering\includegraphics[width=0.40\textwidth]{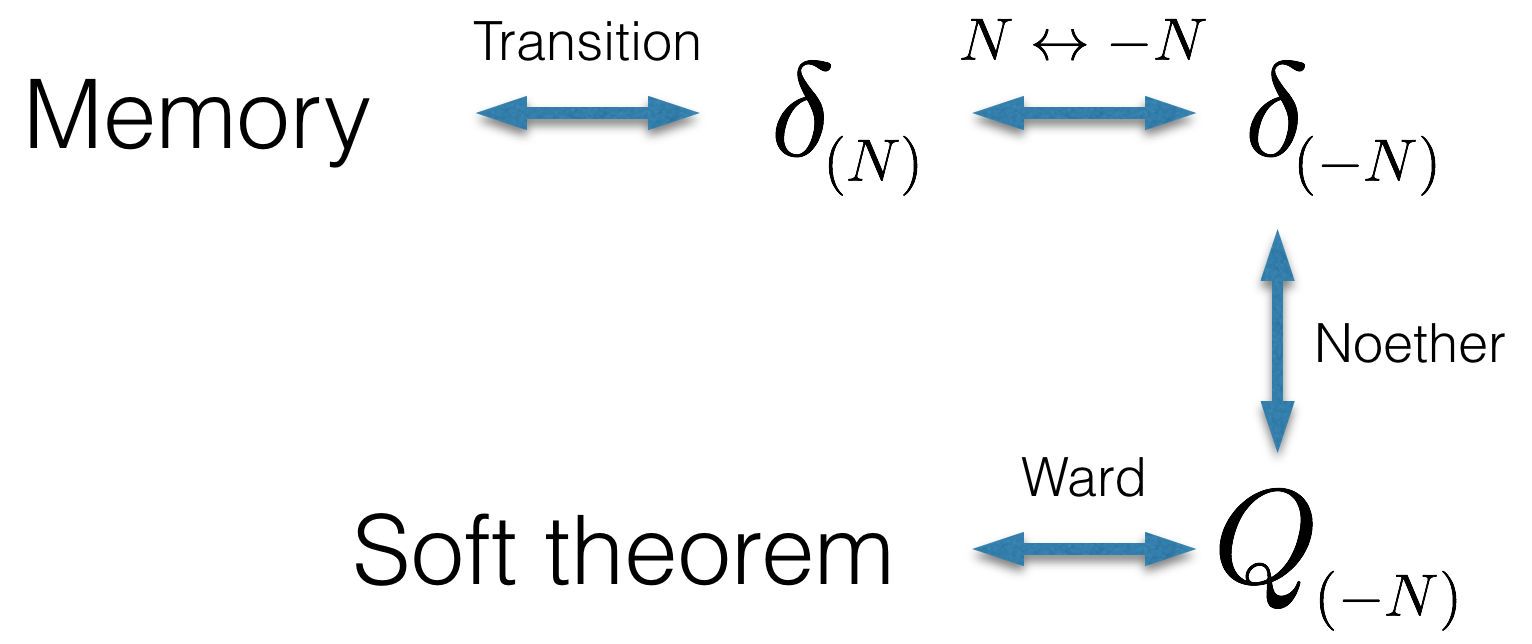}\vspace{-15pt}
\end{figure}

\vspace{7pt}
{\noindent \bf Associated Noether charges.} Gauge transformations are associated with canonical surface charges, which are integrable in non-radiative regions. Overleading gauge transformations are usually discarded since they are not tangent to the phase space (except if one enlarges the phase space and renormalizes the theory, which also leads to overleading memory observables \cite{Compere:2018ylh}). However, as it is  emphasized in several works \cite{Campiglia:2014yka,Campiglia:2016jdj,Campiglia:2016efb,Compere:2017wrj,Compere:2018ylh}, overleading gauge transformations $\delta_{(-N)}$ lead to finite Noether charges $Q_{(-N)}$ \emph{of the standard phase space} that are conserved at spatial infinity, after implementing a subtraction procedure. Given the chain of correspondences presented above, we conclude that \emph{subleading transformations $\delta_{(N)}$ are in one-to-one correspondence with finite conserved Noether charges $Q_{(-N)}$ at spatial infinity that depend upon the subleading orders of the field}. For example, $\xi_{(0)}^{T}$ is associated with the Bondi mass aspect and $\xi_{(1)}^{R}$ is associated with the Bondi angular momentum aspect. In addition, particular linear combinations of the overleading gauge transformations are the multipole symmetries defined in \cite{Compere:2017wrj}, as detailed above, whose associated Noether charges are the mass and current multipole moments \cite{Thorne:1980ru}.

\vspace{10pt}
{\noindent \bf Conserved charges at spatial infinity.} How to define the towers of conserved charges? The simplest answer uses the representation of spatial infinity as $dS_3$ \cite{Ashtekar:1978zz}. The dynamics of gravity can be re-expressed in terms of an infinite tower of symmetric transverse traceless fields $T^{(N)}_{ab}$ on $dS_3$ of the form $(\square + N^2 - 3)T^{(N)}_{ab}=s_{ab}$ where $a,b$ are indices on $dS_3$, $\square$ is the d'Alembertian and $s_{ab}$ are non-linear sources depending on lower subleading fields ($0,\dots N-1$) \cite{1982CMaPh..87...65B} (we ignore here possible logarithmic branches). The initial and final boundary conditions are related by a junction condition, which is necessary to define scattering between $\mathcal I^+$ and $\mathcal I^-$. What is left as independent initial data is a tower of symmetric traceless transverse tensors $C^{(N)}_{ab}(\theta,\phi)$. For $N \geq 0$ such tensors encode 2 arbitrary functions on the sphere as shown e.g. in \cite{Compere:2011db}. I expect that they will precisely encode the conserved charges $Q_{(-N)}$. Such a construction has been performed in electromagnetism \cite{Campiglia:2018dyi}. The non-linearities of gravity are not expected to prevent its generalization.

\vspace{10pt}
{\noindent \bf Soft theorems as Ward identities.} Conserved charges at spatial infinity lead to Ward identities, which are precisely the soft theorems. Their existence go hand-in-hand.  Now, the existence of at least one tower of soft theorems was recently established \cite{Hamada:2018vrw}. More precisely, the soft graviton identities written in \cite{Hamada:2018vrw} can be recognized as the Ward identitites of momentum and current multipole symmetries, which are associated with conserved multipole charges at spatial infinity \cite{Compere:2017wrj}. Such current and momentum multipole symmetries are particular instances of the supertranslations $\xi^T_{(-N)}$ (\ref{xiT}) and superrotations $\xi^R_{(-N)}$ (\ref{xiR})  where $T$ and $R^A$ are harmonics with mode number $N \geq 0$ as detailed above. In contrast, the original derivations of the leading and subleading soft graviton theorems \cite{He:2014laa,Kapec:2014opa} used instead the Ward identities associated with supertranslations $\xi^T_{(0)}$ and superrotations $\xi^R_{(-1)}$ with holomorphic or anti-holomorphic $T$ and $R^A$ as generators. Yet another derivation of the leading, subleading and subsubleading soft theorems was formulated as  a variant of using the Ward identity associated with $\xi^T_{(0)}$, $\xi^T_{(-1)}$ and $\xi^T_{(-2)}$ \cite{Conde:2016rom}. We conclude that the set of Ward identities obtained from the unifying set of generators (\ref{xiT})-(\ref{xiR}) is degenerate in Einstein gravity.


\vspace{10pt}
{\noindent \bf Conclusion.} Memory effects around null infinity are naturally organized in infinite towers, which are mapped by gravitational electric-magnetic duality. Each such memory effect can be  independently generated by suitable matter fields. Such memories are associated with infinite towers of conserved charges at spatial infinity, which also lead to degenerate towers of subleading soft graviton theorems.  The leading supertranslation and leading (novel) superrotations are both associated with a leading displacement memory effect. Consequently,  this suggests the existence of new boundary conditions for asymptotically flat spacetimes that admit the Lorentz group together with mutually commuting supertranslations and superrotations as asymptotic symmetry group.  These considerations are preliminary and many gaps remain to be filled in. A connection with the results of \cite{Duval:2014uva,Duval:2014lpa,Flanagan:2018yzh} remains to be addressed.  In addition, many extensions can be explored such as loop corrections, higher dimensional generalizations,  coupling to matter and alternative gauge theories.

\vspace{10pt}
{\noindent \bf Acknowledgments.} I gratefully thank Adrien Fiorucci, Yegor Korovin, Sabrina Pasterski and Ali Seraj for their comments on the manuscript. I also acknowledge an anonymous referee for his/her constructive comments. I acknowledge support from F.R.S.-FNRS. as research associate.


\providecommand{\href}[2]{#2}\begingroup\raggedright\endgroup

\end{document}